\documentclass[10pt]{wlscirep}
\usepackage{amsmath,graphicx,amssymb,braket,xcolor,subfigure,upgreek}
\usepackage{units}
\usepackage{scalerel,stackengine}
\usepackage{color}

\title{Energy transfer and correlations in cavity-embedded donor-acceptor configurations}

\author[1]{Michael Reitz}
\author[1]{Francesca Mineo}
\author[1,*]{Claudiu Genes}
\affil[1]{Max Planck Institute for the Science of Light, Staudtstra{\ss}e 2,
D-91058 Erlangen, Germany}

\affil[*]{claudiu.genes@mpl.mpg.de}

\begin{abstract}
The rate of energy transfer in donor-acceptor systems can be manipulated via the common interaction with the confined electromagnetic modes of a micro-cavity. We analyze the competition between the near-field short range dipole-dipole energy exchange processes and the cavity mediated long-range interactions in a simplified model consisting of effective two-level quantum emitters that could be relevant for molecules in experiments under cryogenic conditions. We find that free-space collective incoherent interactions, typically associated with sub- and superradiance, can modify the traditional resonant energy transfer scaling with distance. The same holds true for cavity-mediated collective incoherent interactions in a weak-coupling but strong-cooperativity regime. In the strong coupling regime, we elucidate the effect of pumping into cavity polaritons and analytically identify an optimal energy flow regime characterized by equal donor/acceptor Hopfield coefficients in the middle polariton. Finally we quantify the build-up of quantum correlations in the donor-acceptor system via the two-qubit concurrence as a measure of entanglement.
\end{abstract}

\begin{document}
\thispagestyle{empty}

\maketitle

\section*{Introduction}
\begin{figure}[t]
\begin{center}
\includegraphics[width=0.9\textwidth]{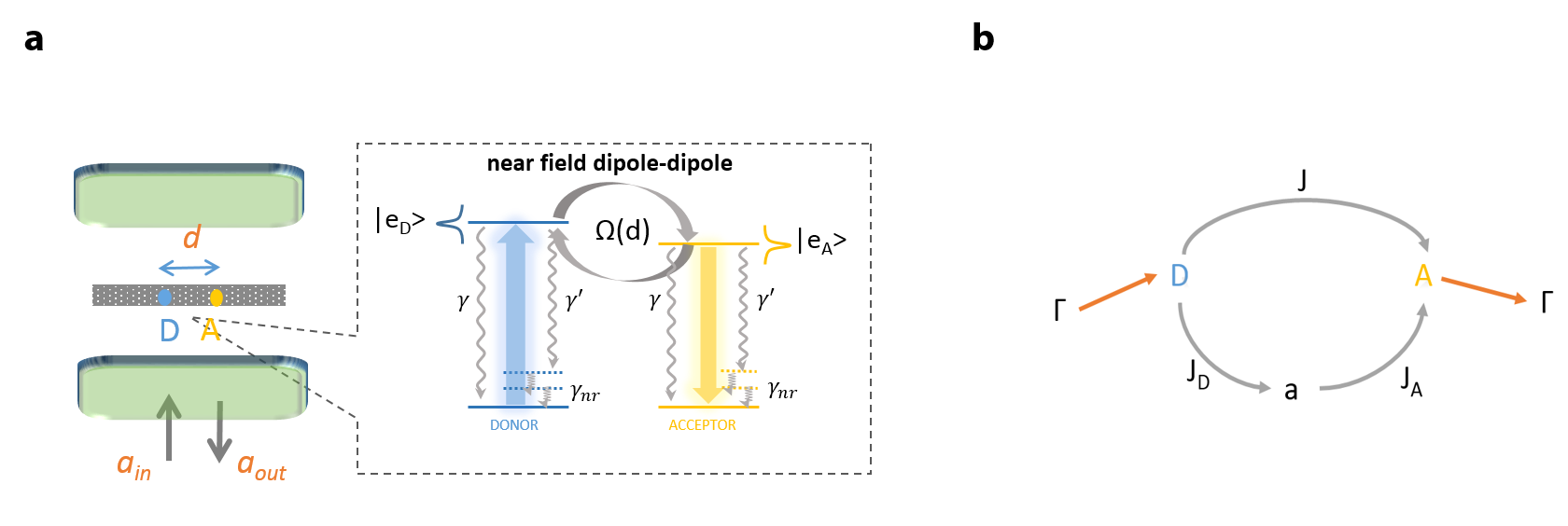}
\end{center}
\caption{\emph{Schematics}. a) Two quantum emitters, in a donor-acceptor-like configuration, are separated by (small) distance $d$ and embedded in an optical microcavity. The relevant internal structure in the inset shows transitions from the excited state to the vibrational manifold of the ground state occurring at rates $\gamma$ (on the zero-phonon line) and $\gamma'$ (for all others). Non-radiative rates $\gamma_{nr}$ characterize dynamics within the vibrational ground state manifold. Beside coupling to the cavity, the molecules can be dipole-dipole coupled with a position dependent rate $\Omega(d)$. b) Representation of the relevant processes including the direct donor-acceptor energy flow rate $J$, mediated by the dipole-dipole coupling and indirect rates $J_A,J_D$ through the cavity mode.}
\label{fig1}
\end{figure}

\indent Resonant energy transfer between light sensitive molecules is a near field effect occurring at separations $d$ much smaller than an optical wavelength, via a virtual photon exchange~\cite{thomas1978energy,scholes2003,loura2012simple}. In a standard picture, one considers a scenario where the zero-phonon electronic vibrational excited state of a light-absorbing molecule (the donor) lies far above the zero-phonon electronic vibrational excited state of the energy accepting molecule (the acceptor). In light-harvesting systems, under high-temperature conditions, energy transfer between incoherently pumped donor states and the excited vibrational manifold of the acceptor takes place mediated by the inherent dipole-dipole interactions (scaling as $1/d^3$). The process is unidirectional as a resonant virtual exchange of photons is followed by a high-rate incoherent non-radiative relaxation to the zero-phonon excited state of the acceptor. This quantifies a resonant F\"{o}rster energy transfer rate proportional to $1/d^6$, that can be seen as arising from the overlap integral between the emission spectrum of the donor and the absorption spectrum of the acceptor~\cite{thomas1978energy}.\\
Controlled experiments under cryogenic conditions provide an alternative platform for the detailed description and manipulation of such effects~\cite{wang2017coherent}. We focus in this manuscript on scenarios where coherent or incoherent pumping of the zero-phonon line of a donor molecule is followed by energy exchange with an acceptor controllably detuned with frequency differences much smaller than vibrational separations. This justifies a three level system model for interacting quantum emitters where two vibrational states of the ground state vibrational manifold are considered, coupled by non-radiative rates much larger than the radiative ones. Our aim is twofold: i) to investigate the effect of collective dynamics of vacuum-coupled quantum emitters (exhibiting subradiance and superradiance) on the energy flow and ii) to analyze the role of cavity-mediated processes in the energy transfer process.\\
The investigations are triggered by recent experimental results~\cite{zhong2016non,zhong2017energy} carried out in microcavity settings albeit in a different regime; in these experiments organic molecules are collectively and strongly coupled to highly confined optical modes either in  microcavities or near plasmonic structures. It has been unveiled that some material properties involving the interaction of light and matter can be strongly modified by the reach of the strong coupling regime as occurring for example in cavity quantum electrodynamics. Examples are: charge and energy transport enhancement in organic semiconductors~\cite{orgiu2015conductivity,feist2015extraordinary,schachenmayer2015cavity,hagenmuller2017cavity,hagenmuller2018fermions}, cavity quantum chemistry showing reaction rate slow-downs or modifications of molecular bond lengths~\cite{shalabney2015coherent,george2015liquid,galego2016suppressing,galego2017many,flick2017atoms,flick2017cavity}, etc. The process of F\"{o}rster energy transfer has been tackled as well in experiments showing a vacuum-induced enhancement of the transfer for donor-acceptor molecular layers inside an optical microcavity~\cite{coles2013polariton,zhong2016non,zhong2017energy}. While initial experiments have shown a competition between free space and cavity induced effects~\cite{zhong2016non}, a follow-up treatment involving spatially distant donor-acceptor layers has revealed a purely cavity induced energy transfer of high efficiency~\cite{zhong2017energy}. Theoretical considerations suggest that non-radiative processes that can couple light-matter polaritons especially in a dissipative fashion can play a role in the modification of the energy transfer~\cite{feist2017long,vidal2018organic,joel2018theory}. The fundamental mechanism of energy transfer mediated by strongly confined light modes has also been experimentally tackled in nano-optical systems such as between two nanoemitters attached to microspheres~\cite{gotzinger2006controlled}. Moreover, high finesse microcavities promise the reach of strong coupling at the single particle level and allow for a purely microscopic and controllable study of energy transfer between quantum emitter-like systems~\cite{cano2010resonance}. \\
In the free space scenario, we quantify the energy flow $J$ [see Fig~\ref{fig1}] as a function of particle-particle distance and find that the inherent collective decay accompanying dipole-dipole interactions can lead to enhancement of the energy flow and even its sign reversal. Such effects are strictly brought on by the access to subradiant states which allows increased population accumulation for constant pump strength. We also analyze energy flow and donor-acceptor entanglement (quantified via the quantum concurrence measure) to reveal that generally the two quantities are not connected and even mutually exclusive under certain conditions. As our analytical treatment is restricted by the assumption of weak excitations, one can show that the set of equations we analyze are perfectly classical. A classical approach therefore suffices to characterize the energy transfer process~\cite{Duque2015quantumclassical}; however, this forbids entanglement which occurs when departing from the classical description including higher particle-particle correlations (non-factorizable products of operators). While concurrence is present in the system, the occurrence of entanglement does not affect in any way the energy transfer rates either in free space or cavity settings. We then extend results obtained in free space to reveal effects of cavity subradiance and superradiance onto energy flow under strong cooperativity but weak coupling conditions. Under such conditions, a leaky cavity field can enlarge the emission and absorption spectra of the donor and acceptor and lead to an increase in the direct rate $J$. In the opposite case where a good cavity leads to the occurrence of polaritons (strong coupling conditions), we analyze optimal strategies for enhancing the long-distance cavity mediated transfer $J_A$. We analytically derive the positioning of the cavity resonance with respect to the donor/acceptor resonances for optimal transfer: this occurs at the intersection of donor/acceptor Hopfield coefficients in the middle cavity polariton.\\

\noindent \textbf{Model} - We consider two types of molecules (donor-type D and acceptor-type A) with states $\ket{g_{D,A}}$ and $\ket{e_{D,A}}$, standing for ground and excited states respectively. The electronic transition frequencies are $\omega_{D,A}$ with $\Delta=\omega_D-\omega_A$. The relaxation rates are $\gamma$ (for the zero-phonon line) and $\gamma'$ (for any other radiative transitions from the excited state) and $\gamma_{nr}$ for quick phononic relaxation of the ground state vibrations. For small separations $d\ll\lambda_{D,A}$, where $\lambda_{D,A}=2\pi c/\omega_{D,A}$, two molecules interact via near-field effects where virtual photon emission-absorption processes amount to both coherent and incoherent interactions~\cite{ficek1987quantum}. The coherent interaction consists of an effective dipole-dipole exchange quantified by the energy shift $\Omega$ (see Methods for exact expression) that takes into account the separation and the orientation of dipoles with respect to the interparticle axis~\cite{ficek1987quantum}. The incoherent interaction is quantified by a mutual decay rate $\bar {\gamma}$ responsible for sub-and superradiant behavior of symmetric and asymmetric hybrid states of the two-molecule system~\cite{ficek1987quantum}. Both $\Omega$ and $\bar {\gamma}$ are proportional to the zero-phonon line radiative rate $\gamma$ such as at close separations, $\bar{\gamma}$ saturates to $\gamma$. The formalism we follow is based on solving the master equation $\partial_t\rho=-i[H,\rho]+\mathcal{L}[\rho]$ for the density operator of the full donor-acceptor-cavity system $\rho$ under the action of coherent $H$ and incoherent $\mathcal{L}[\rho]$ processes. Incoherent processes are included as Lindblad terms $\mathcal{L}_{\Gamma_O} [\rho]=\Gamma_O D_O[\rho]= \Gamma_O \left[O \rho O^{\dagger}-\left\{O^{\dagger} O,\rho \right\}/2\right]$ and are characterized by rates $\Gamma_O$ for each collapse operator $O$ (the brackets $\left\{.,.\right\}$ stand for anticommutation). The incoherent terms are: i) donor/acceptor population decay assumed for simplicity at rate $\gamma_{D,A}$ with collapse operators $\sigma_{D,A}$, ii) donor/acceptor dephasing owing to coupling to bulk phonons for example at rate $\gamma_\phi$ and with operators $\sigma^z_{D,A}$, iii) cavity decay rate at rate $\kappa$ with collapse operator $a$. Two extra incoherent processes are modelled as pump (acting on the donor with rate $\Gamma$ and operator $\sigma_D^{\dagger}$) and drain (acting on the acceptor with rate $\Gamma$ and operator $\sigma_{A}$). The pumping could stem either from driving with large-bandwidth incoherent light or from coherently driving to a manifold of excited vibrational donor states followed by non-radiative relaxation to the $\ket{e_{D}}$ state of interest. In reality, the physics of light-harvesting systems under illumination by incoherent natural thermal light is quite complex as revealed in Ref.~\cite{Pachon2017incoherent}. We therefore assume a more simple model realizable under experimental conditions such as in Ref.~\cite{wang2017coherent} where coherent laser excitation on an additional excited vibrational level is followed by non-radiative decay to the zero-vibrations excited level. In the Methods section, we provide a simple adiabatic elimination method showing the emergence of an effective Lindblad damping term, that can also be viewed as decay from the ground to the excited level. We also compare results with a coherent drive model where we directly address the donor with a monochromatic laser $\omega_L$ away from the acceptor frequency and pump strength $\eta$. In such a case the pumping process is simply included in the Hamiltonian (in a frame rotating at the acceptor): $H_d=\eta (\sigma_D e^{i\omega_L t}+\sigma^{\dagger}_D e^{-i\omega_L t})$. Near field effects contribute as well which are casted in a non-standard form as $\mathcal{L}_{\bar{\gamma}} [\rho]=\bar{\gamma} \left[\sigma_D \rho \sigma^{\dagger}_A+\sigma_A \rho \sigma^{\dagger}_D-\left\{\sigma^{\dagger}_A\sigma_D+\sigma^{\dagger}_D\sigma_A,\rho \right\}/2\right]$. \noindent As discussed in detail in Methods, the presence of the mutual decay rate stemming from near field effects leads to different effective decay rates for symmetric/asymmetric states $1/\sqrt{2}(|e_Dg_A\rangle \pm |g_De_A\rangle )$ at superradiant/subradiant rates $\gamma\pm\bar{\gamma}$. In the so-called Dicke limit of $d=0$, $\bar{\gamma}\rightarrow \gamma$ such that the subradiant state is theoretically infinitely long lived. The coherent processes can be summed up in the total Hamiltonian
\begin{equation}
  H = \Delta \sigma^{\dagger}_D \sigma_D + \delta a^{\dagger}a+\Omega (\sigma^{\dagger}_D \sigma_A+\sigma^{\dagger}_A \sigma_D)+g_D (\sigma^{\dagger}_D a+a^{\dagger} \sigma_D)+g_A (\sigma^{\dagger}_A a+a^{\dagger} \sigma_A),
\end{equation}
written in a frame rotating at the acceptor's frequency such that $\delta=\omega-\omega_A$ (with $\omega$ the cavity resonance). The third term represents the near-field exchange while the last two terms are the cavity Jaynes-Cummings interactions with field-donor and field-acceptor couplings $g_D$ and $g_A$. For experiments involving single or a few molecules in cavities decaying at rates around GHz~\cite{wang2017coherent} the requirement for strong coupling $g_{D,A}>\kappa$ is fulfilled long before entering the ultra-strong coupling regime characterized by $g_{D,A}\approx \omega_{D,A}$. This is the focus of this manuscript. On the other hand, experiments with many organic molecules in bad cavities have decay rates of the order of tens to hundreds THz~\cite{zhong2016non,zhong2017energy}, such that the strong coupling regime is very close to the ultra-strong coupling regime. Just as in the case of experiments involving superconducting artificial atoms coupled to on-chip cavities, the simple Jaynes-Cummings model is then insufficient to properly describe the dynamics~\cite{niemczyk2010ultrastrong}. \\

The efficiency of the energy transfer can be quantified by the changes in the acceptor population owing to the coupling to the donor system. A relevant quantity characterizing this process is the direct steady state energy flow rate $J = 2 \Omega \Im \braket{\sigma_D \sigma^{\dagger}_A}$ representing the net energy transfer between donor and acceptor. The other relevant quantities similarly defined are $J_D = 2 g_D \Im \braket{\sigma_D a^{\dagger}}$ and $J_A = 2 g_A \Im \braket{a \sigma^{\dagger}_A}$ characterizing jump processes from the donor to the cavity mode and from the cavity to the acceptor, respectively. In the simplest case of two-level emitters independently decaying and coupled to a single cavity mode, it is instructive to write equations of motion for two-operator correlations (see Methods), and to derive steady state rate equations showing migration of population from the donor to the acceptor:
\begin{eqnarray}
   \dot{p}_D &=&\Gamma-(\Gamma+\gamma)p_D-J-J_D,\\
   \dot{p}_A &=&-(\Gamma+\gamma)p_A+J+J_A.
\end{eqnarray}
Generally, the source terms in the acceptor equation cannot be cast as a rate $\sim p_A$ as it is the case for the realistic case of F\"orster energy transfer in complex systems characterized by many vibrational levels. In that case, transfer to the vibrational manifold of the acceptor followed by quick decay via non-radiative processes insures the irreversability of the energy transfer process and can be characterized by an energy transfer rate. Our equations treat a reversible problem instead, where both directions donor-acceptor and acceptor-donor are allowed, which will eventually lead to new effects such as energy flow direction change under subradiant conditions. To avoid confusion we therefore denote the derived quantities $J$ and $J_A$ as energy flow rates in the following analysis.

\section*{Results}

\subsection*{Free space resonant energy transfer including collective effects and dephasing}
In free space, the only energy flow mechanism is provided by the direct donor-acceptor interaction. We proceed in analyzing the dependence of the energy flow rate $J$ on the separation $d$ between two similar quantum emitters in various regimes characterized by the presence of extra dissipation channels, collective decay and dephasing. To this end we solve the set of equations connecting energy flows to populations (Eq.~\ref{complete}-\ref{end} in the Methods section) in the absence of cavity couplings (setting $g_A$ and $g_D$ to zero) and in steady state. A general fully analytical result can be obtained relating the normalized energy flow
\begin{equation}
J/\Gamma=\frac{2\Omega^2\left[\frac{\gamma+\gamma'+2\gamma_\phi}{(\gamma+\gamma')}-\frac{\bar{\gamma}^2}{(\gamma+\gamma')^2}\right]+\frac{\bar{\gamma}\Delta\Omega}{(\gamma+\gamma')}}{\Delta^2+4\Omega^2\frac{\gamma+\gamma'+2\gamma_\phi}{(\gamma+\gamma')}+(\gamma+\gamma'+2\gamma_\phi)^2-{\bar{\gamma}^2}\left[\frac{4\Omega^2}{(\gamma+\gamma')^2}+\frac{\gamma+\gamma'+2\gamma_\phi}{(\gamma+\gamma')}\right]}
\label{total}
\end{equation}
to quantities such as additional population dissipation rates $\gamma'$, environmentally induced dephasing rate $\gamma_{\phi}$ and mutual decay rate $\bar{\gamma}$. The multiple level scheme containing decay from the excited state to states non-radiatively coupled to the ground state leads to a renormalization of the effective two level system decay rate from $\gamma$ to $\gamma_{tot}=\gamma+\gamma'$. The general expression above is a linearized form that assumes weak pumping such that the energy flow properties of the system are pumping independent besides the expected linear behavior. Notice that this assumption implies a regime of validity where $\Gamma\ll\gamma_{tot}-\bar{\gamma}$ outside of which the derived quantities do not reach steady state and which is straightforward to derive by applying stability conditions to the evolution matrix. Notice also that $\bar{\gamma}$ is a mutual decay rate associated with the direct excited-ground transition rate $\gamma$, therefore having an upper bound $\gamma$ reached at very small distances. In treating the ideal case of pure two-level emitters (where $\gamma'=0$) special care has to be given to the small distances cases where $\gamma_{tot}-\bar{\gamma}$ tends to zero. \\
Setting $\gamma_{\phi}=0$ and $\bar{\gamma}=0$ we recover an expected simple expression
\begin{equation}
J/\Gamma=\frac{2\Omega^2} {\Delta^2+4\Omega^2+\gamma_{tot}^2},
\end{equation}
revealing the $d^{-6}$ scaling (stemming from the near field behavior of the $\Omega^2$ term - see Methods) of the energy flow rate with particle-particle separation for $\Delta\gg |\Omega|$ and a maximum energy flow rate of $J/\Gamma=1/2$ in a resonant regime characterized by vanishing energy separation $\Delta=0$ and $|\Omega|\gg\gamma_{tot}$. The denominator can simply be seen as emerging from the overlap of two Lorentzians of widths $\gamma_{tot}$ characterizing the emission spectrum of the donor and the absorption spectrum of the acceptor. A straightforward generalization, that we will find useful in the next section, can be performed for distinct emitters with $\gamma_A\neq \gamma_D$. Neglecting extra dissipation channels we arrive at the following expression
\begin{equation}
J/\Gamma=\frac{4\gamma_A(\gamma_A+\gamma_D)\Omega^2}{\gamma_A\gamma_D\left[4\Delta^2+(\gamma_A+\gamma_D)^2\right]+4(\gamma_A+\gamma_D)^2\Omega^2},
\label{JADdifferent}
\end{equation}
\begin{figure*}[t]
\begin{center}
\includegraphics[width=0.8\textwidth]{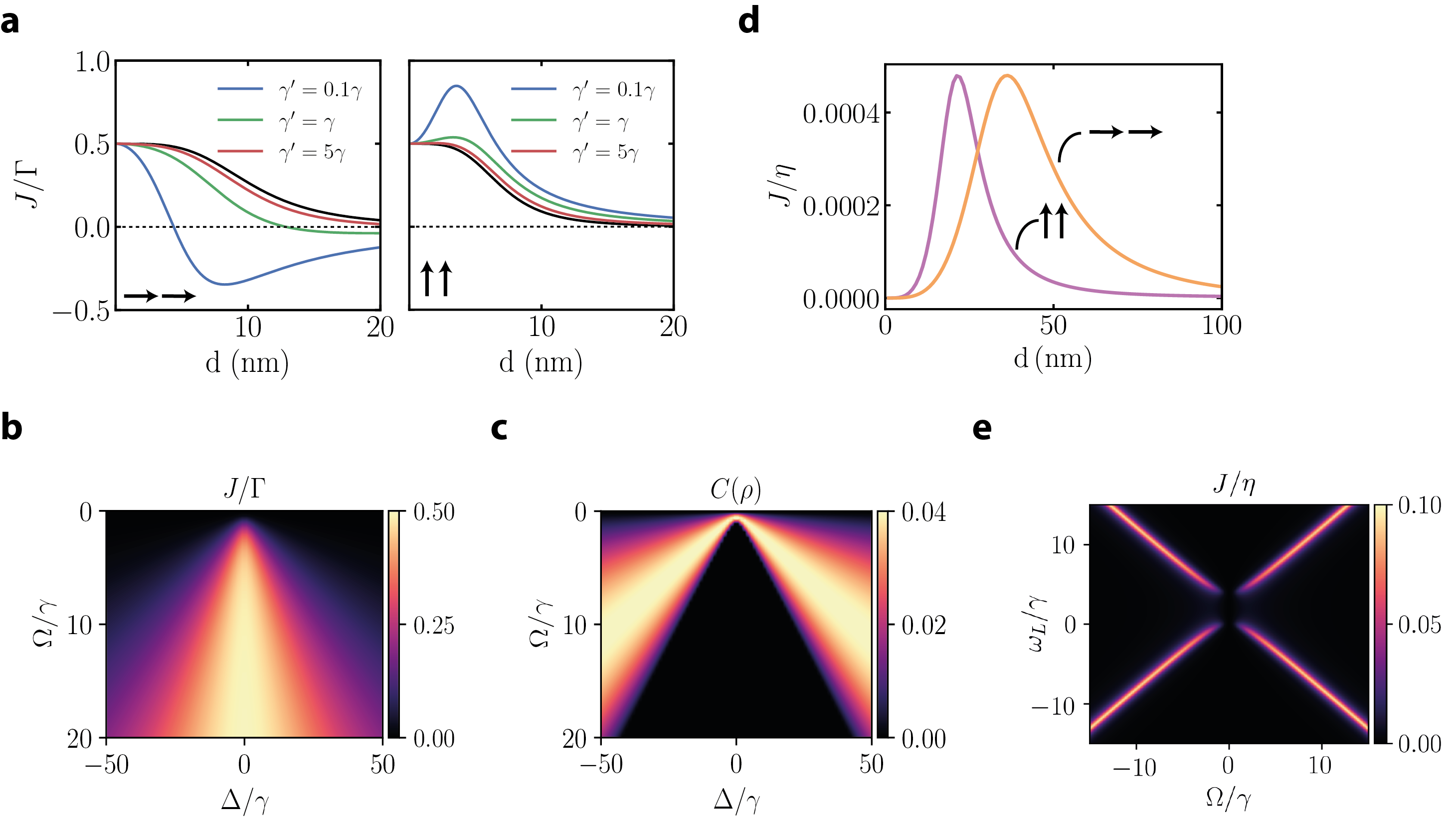}
\end{center}
\caption{\emph{Free space results}.  a) Normalized energy flow $J/\Gamma$ versus donor-acceptor separation for the parallel case ($\mathbf{\mu}_{A,D} \parallel \mathbf{d}$, left) and for the perpendicular case ($\boldsymbol{\mu}_{A,D} \perp \mathbf{d}$, right) for a detuning $\Delta=200\gamma$. The black lines represent the result without collective decay $\bar{\gamma}=0$. Including collective decay leads to a modified behavior which is quite significant in the presence of small additional radiative decay ($\gamma'=0.1 \gamma$ - blue curve) but is eventually washed out with increasing $\gamma'$ (green and red curves). b) Energy flow rate $J/\Gamma$ as a function of detuning $\Delta$ and dipole-dipole coupling strength $\Omega$  and c) comparison to steady state concurrence $C(\rho)$ (without collective decay). d) Energy flow $J/\eta$ for coherent driving as function of distance for $\boldsymbol{\mu}_{A,D} \parallel \mathbf{d}$ (orange) and $\boldsymbol{\mu}_{A,D} \perp \mathbf{d}$ (purple) for $\eta=0.1\gamma$, $\Delta=200\gamma$, $\omega_L=\Delta$. e) Energy flow $J/\eta$ for coherent driving versus laser frequency $\omega_L$ and dipole-dipole coupling $\Omega$ for $\eta=0.1\gamma$, $\Delta=4\gamma$.}
\label{fig2}
\end{figure*}
again obtainable via simple arguments as an overlap emission/absorpion spectra for different species. Notice that an asymmetry is now included as the main tuning parameter is now $\gamma_A$; a narrow acceptor absorption window can inhibit energy flow while for $\gamma_A\gg\gamma_D$ and small separations $J\rightarrow\Gamma$.\\
Surprising results are predicted by Eq.~\ref{total} in the regime where the intrinsic collective decay is considerable ($\bar \gamma$ becomes of the order of $\gamma$ for $d \ll \lambda$). We plot in Fig.~\ref{fig2} results for the scaling of the energy flow for different dipole alignment configurations and for a transition associated wavelength of $500$~nm. Figure~\ref{fig2}a shows that subradiant/superradiant behavior is linked to an uncharacteristic scaling of the energy flow at small distances and even a change in the direction of the flow from the acceptor to the donor. The effect stems from pumping into subradiant states. A simple argument for such behavior can be understood from the following proportionality: from Eqs.~\ref{complete} one can show that $J\propto(p_D-p_A)$, i.e. the energy flow magnitude and direction are proportional to the population imbalance in steady state. While pumping an independently decaying system, both populations would be of the order of $\Gamma/(\gamma+\gamma'+\Gamma)$, pumping into collective states, one of them characterized by a subradiant rate $\gamma+\gamma'-\bar \gamma$ can give rise to a more efficient population accumulation proportional to $\Gamma/(\Gamma+\gamma+\gamma'-\bar \gamma)$. The effect is an increased effective energy flow $\Gamma \ll J$. However, close inspection of the equations of motion shows that a physically necessary upper bound $ J\ll\gamma$ occurs.\\
The efficiency of pumping into subradiant states is however easily blurred under realistic scenarios by the presence of extra decay channels ($\gamma'\neq0$) or dephasing ($\gamma_{\phi}\neq0$). The analysis of the connection between entanglement and efficient flow as shown in Figure~\ref{fig2}b,c. As a characterization of entanglement we make use of concurrence (monotonously increasing from 0 for separable states to unity for maximally entangled Bell states) as defined in Methods. As apparent from a scan of detuning around regions of efficient energy flow, the concurrence is maximized around the borders of efficient energy flow. While the full numerical solution can show large levels of concurrence, we restrict this treatment on the low excitation regime where the validity of our linearized treatment is insured. The alternative approach characterized by a coherent drive of the donor indicates the following analytical result ($\bar{\gamma},\gamma',\gamma_\phi=0$) for the energy flow rate as a function of pump strength $\eta$ and of the detuning between the laser frequency and the acceptor $\omega_L$
\begin{equation}
J=\frac{16 \gamma  \eta ^2 \Omega ^2}{\left(\gamma ^2+4  \omega^2_L\right) \left[\gamma ^2+4
  \left(\left(\Delta-\omega_L\right)^2+2 \eta
   ^2\right)\right]+8 \Omega ^2 \left[\gamma ^2+4 \omega_L
   \left(\Delta-\omega_L\right)\right]+16 \Omega ^4}.
\end{equation}
The behavior of the energy flow rate with respect to the interparticle distance is illustrated in Fig.~\ref{fig2}d, for both dipoles parallel or orthogonal to $\mathbf{d}$. The direct discrepancy with respect to the incoherent pump case is the vanishing rate at small distances. As the pump is monochromatic, in the regime of strongly interacting emitters (small separations) the resonance condition is shifted away from the bare donor frequency. The physics is completely elucidated in  Fig.~\ref{fig2}e where we employ an additional scan of the laser drive frequency to reveal the shifting of the resonances to closely follow the linear dependence  of the symmetric/antisymmetric resonance energies with $\Omega$.

\subsection*{Cavity mediated energy transfer}
Let us now consider additional energy transfer channels brought on by the presence of a common confined mode of a microcavity (see Fig.~\ref{fig1}b). The analysis is quite complex as various regimes can occur characterized by the ratios $g_{D,A}/\kappa$ as well as $C_{D,A}=4 g_{D,A}^2/(\kappa\gamma)$ (which we denote cooperativity and is proportional to the Purcell factor). We mainly distinguish between weak cooperativity (and also weak coupling) regimes with $\gamma\ll g_{D,A}\ll \kappa$ and $C_{D,A}<1$, strong cooperativity but still weak coupling regime with $\gamma\ll g_{D,A}\ll \kappa$ but $C_{D,A}>1$ and strong coupling regime with $g_{D,A}/\kappa \gg 1$. In the first two regimes the effect of a bad cavity is mainly in enhancing the dissipation rate of the donor or acceptor (depending on the choice of $\delta$), which can lead to an increased overlap between the emission spectrum of the donor and the absorption spectrum of the acceptor. In the strong coupling regime, polariton dynamics occurs as the emitters are hybridized by the cavity field. \\

\begin{figure*}[t]
\begin{center}
\includegraphics[width=0.7\textwidth]{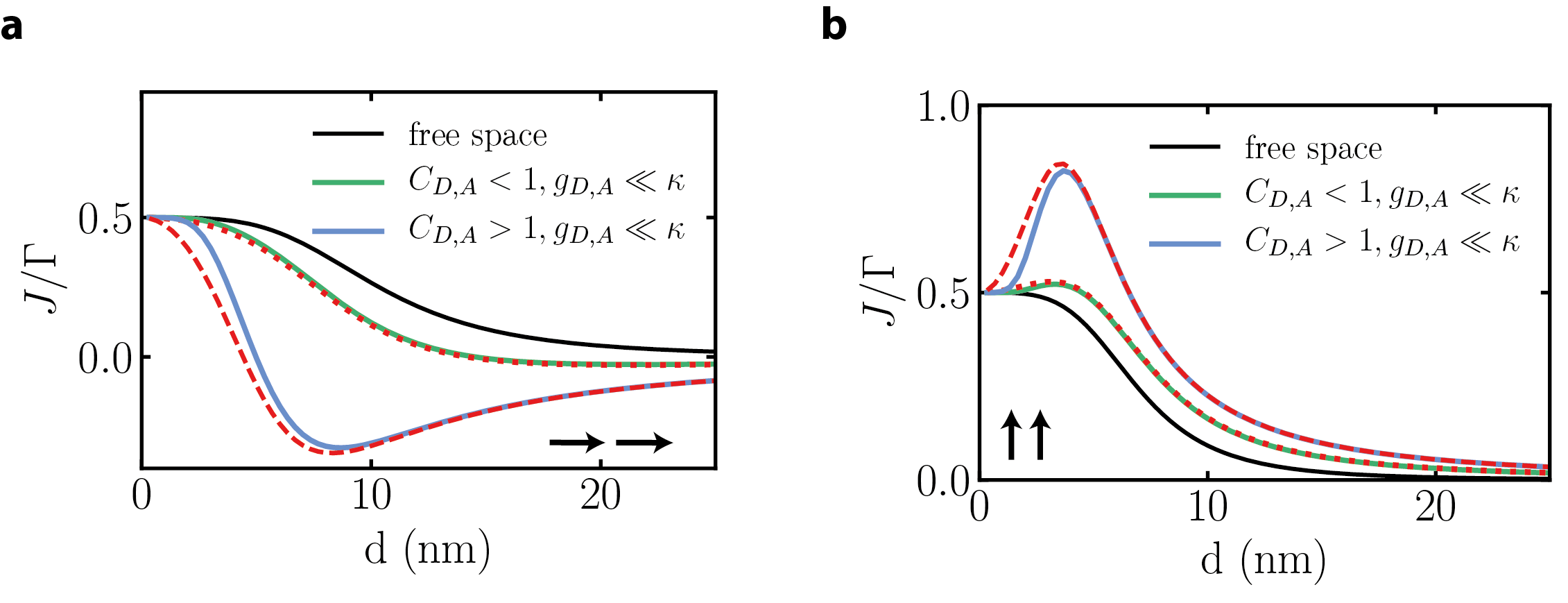}
\end{center}
\caption{\emph{Strong cooperativity results}. Modification of free space energy transfer (black lines) by cavity induced collective decay for a) dipole orientation parallel to $\mathbf{d}$ and b) dipole orientation perpendicular to $\mathbf{d}$, for $\kappa=2000\gamma$. The green and blue lines correspond to exact numerical results with $g_A, g_D=20\gamma$ (green) and $g_A, g_D=70\gamma$ (blue). The dashed red lines correspond to the analytical result from {Eq.~\ref{cooperativitycavity}}.}
\label{fig3}
\end{figure*}

\noindent \textit{Strong cooperativity regime} - In the bad cavity regime, analytical simplifications can be employed when conditions are met for the adiabatic elimination of the cavity mode. Assuming either $\kappa\gg g_A,g_D$ we can find an adiabatic solution for the intracavity field:
\begin{equation}
\label{adelim}
\braket{a(t)}= -\frac{ig_D}{\kappa/2+i(\delta-\Delta)} \braket{\sigma_D(t)}-\frac{ig_A}{\kappa/2+i\delta} \braket{\sigma_A(t)},
\end{equation}
that allows us to recast the equations of motion between the donor and acceptor as including additional couplings owing to the common interaction with the cavity mode. The noise terms are not explicitly written as the usual fluctuation-dissipation theorem insures that they are consistent with the derived decay rates. An effective coherent and incoherent evolution for the donor-acceptor system can be derived with the main result that cavity-modified decay rates as well as a mutual decay rate occur:
\begin{subequations}\label{effdecay}
\begin{eqnarray}
\tilde{\gamma}_A=\gamma_A+\frac{g_A^2\kappa}{\left(\kappa/2\right)^2+\delta^2}\approx \gamma_A\left[1+C_A(\delta)\right]\\
\tilde{\gamma}_D=\gamma_D+\frac{g_D^2\kappa}{\left(\kappa/2\right)^2+(\delta-\Delta)^2}\approx \gamma_D \left[1+C_D(\delta)\right] \\
\tilde{\gamma}_{AD}\approx \frac{g_D g_A\kappa}{\left(\kappa/2\right)^2}=\sqrt{\gamma_A\gamma_D}\sqrt{C_A(0) C_D(\Delta)}
\end{eqnarray}
\end{subequations}
\
where $C_A(0)=4g_A^2/(\kappa\gamma_A)$, $C_D(\Delta)=4 g_D^2 / (\kappa\gamma_D)$ are the maximum cooperativities of acceptor and donor, respectively. An expected result is the Purcell enhancement of the two independent emitter decay rates. However, the crucial term is the occurrence of a cavity-mediated mutual decay rate proportional to the geometrical average of the two cooperativities. The solution is straightforward as it can be easily written down by identification with the expression in Eq.~\ref{total} and under the assumption that $\tilde{\gamma}_A=\tilde{\gamma}_D=\tilde{\gamma}$:
\begin{equation}
J/\Gamma=\frac{\Omega\left(2\tilde{\gamma}^2\Omega-2\tilde{\gamma}_{AD}^2\Omega+\tilde{\gamma}\tilde{\gamma}_{AD}\Delta\right)}{\tilde{\gamma}^4-4\tilde{\gamma}_{AD}^2\Omega^2+\tilde{\gamma}^2\left(\Delta^2+4\Omega^2-\tilde{\gamma}_{AD}^2\right)}.
\label{cooperativitycavity}
\end{equation}
Comparison of the above analytical expression with exact numerical simulations (see Fig.~\ref{fig3}) show almost perfect agreement, validating that cavity-mediated collective subradiant behavior leads to the same effect as previously unveiled in the free-space case. The effect of the cavity in this regime is to 'dress' the free space interaction $J$ and not in providing an extra channel independent of emitter-separation (the corresponding $J_A$ is close to zero). This is in stark contrast to the situation tackled in the next subsection under strong coupling conditions: there, even for large distances characterized by vanishing $J$, energy transfer is large stemming solely from the $J_A$ contribution.\\

\begin{figure*}[t]
\begin{center}
\includegraphics[width=0.77\textwidth]{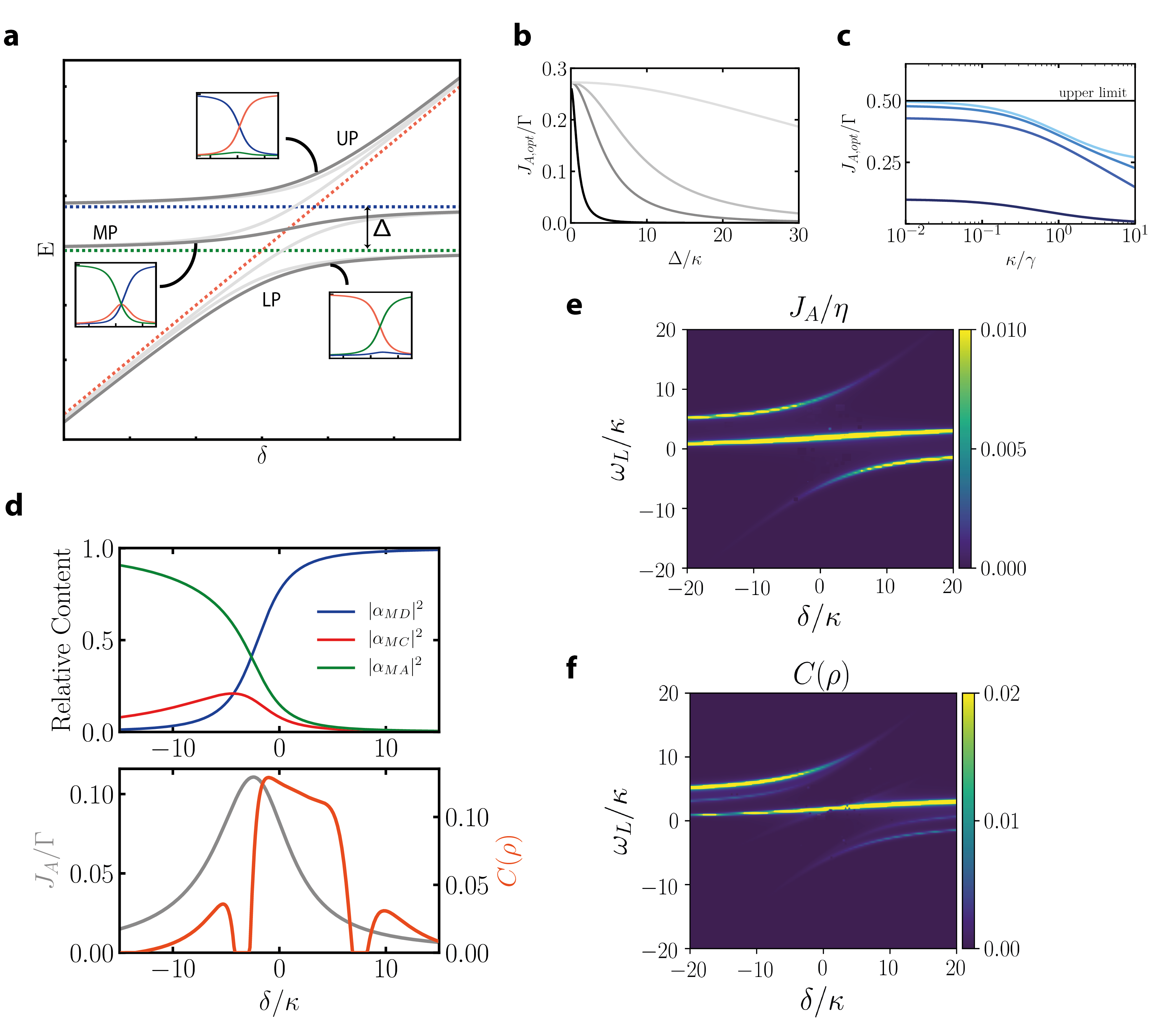}
\end{center}
\caption{\emph{Strong coupling regime}.  a) The energy of the polaritons of the total system (dark grey) as well as individual donor/acceptor polaritons (light grey) as a function of the cavity frequency (detuning $\delta$ with respect to the acceptor's resonance). Dashed lines show the energies of the bare (uncoupled) system. The insets show the donor (blue), cavity (red) and acceptor (green) Hopfield coefficients for each polariton. b) and c)  Energy flow $J_{A, opt}$ obtained at the optimal working points for varying donor-acceptor detuning  and varying cavity decay rate (from dark to light colors: $g_A,g_D=10\gamma, 50\gamma, 100\gamma, 500\gamma$). d) Upper panel:  Hopfield coefficients for the middle polariton versus energy flow rate $J_A/\Gamma$ (grey) and steady state concurrence $C(\rho)$ (orange) (lower panel) for $g_A=50\gamma$, $g_D=10\gamma$, $\kappa=10\gamma$, $\Delta=40\gamma$. e) and f) Results for coherent driving: $J_A/\eta$ and steady state concurrence $C(\rho)$ versus both cavity frequency $\delta$ and laser frequency $\omega_L$ for $g_A=g_D=50\gamma$, $\kappa=10\gamma$, $\Delta=40\gamma$. }
\label{fig4}
\end{figure*}
\noindent \textit{Strong coupling regime} - In the strong coupling regime the coherent processes dominate over dissipation rates. We focus our analysis on distant emitters ($\Omega=0$) to analyze the effect of the cavity mode on transfer in regimes where no direct transfer is expected. In the uncoupled basis of donor-cavity and acceptor-cavity, donor-like polaritons occur at $2\mathcal{E}^\pm_D=\Delta+\delta\pm\sqrt{4 g_D^2+(\Delta-\delta)^2}$ and acceptor-like polaritons occur at $2\mathcal{E}^\pm_A=\delta\pm\sqrt{4 g_D^2+\delta^2}$ (see Fig.~\ref{fig4}a). However, the correct description of the system which allows an analytical understanding is obtained by considering the proper basis for the system made up by three new eigenstates which we denote as upper, middle and lower polariton. These are obtained from diagonalizing the Hamiltonian $H$ matrix:
\begin{equation}
H=\left(
  \begin{array}{ccc}
    \Delta & g_D & 0 \\
    g_D & \delta & g_A \\
    0 & g_A & 0 \\
  \end{array}
\right).
\end{equation}

\noindent While analytical expressions are not readily available, we will write in the single excitation regime symbolically the new basis: $ |P\rangle = \alpha_{PD} |D\rangle + \alpha_{PC} |C\rangle + \alpha_{PA} |A\rangle $ with $P$ standing for $U,M,L$ (upper/middle/lower) and $|\alpha_{PD,PC,PA}|^2$ denote the Hopfield coefficient of each polariton (the contribution of donor/cavity/acceptor state to the respective polariton). The incoherent Lindblad term that populates the donor can then be written in the coupled basis as a pump into all three polariton populations and six polariton-polariton coherences. Optimal donor-acceptor energy flow rate occurs then when the middle polariton contains equal content of donor and acceptor components (minimizing the cavity Hopfield coefficient at the same time). While the polariton expressions are not analytically expressable, the intersection of the donor and acceptor Hopfield coefficients can be cast in a very simple form:
\begin{equation}
\delta_{opt}=\frac{g_D^2-g_A^2}{\Delta}+\frac{g_A\Delta}{g_A+g_D}
\end{equation}
Figure~\ref{fig4} shows numerical evidence of the polariton physics described above. A sketch (arbitrary units) is shown in Fig.~\ref{fig4}a, where the energy of the polaritons is plotted for varying cavity-acceptor detuning $\delta$ with insets showing the corresponding Hopfield coefficients. Figure~\ref{fig4}d illustrates that the reach of an optimal energy flow is at the middle polariton Hopfield intersection where donor-acceptor coefficients are equal. Under optimal conditions (fixing the cavity frequency at $\omega_c=\omega_A+\delta_{opt}$ we then perform a numerical analysis (Figs.~\ref{fig4}b,c) of the energy flow rate $J_A$ with varying controllable parameters such as $\Delta$ and $\kappa$. Figure~\ref{fig4}c shows the transition from bad to good cavity regimes (weak to strong coupling) exhibiting a transition region around the onset of strong coupling condition.
For the more clear understanding of the hybrid resonances in the system, we employ again a coherent drive scheme which allows an analytical expression for the energy rate:
\begin{equation}
J_A= \frac{g_A^2 g_D^2 \gamma \eta^2}{\left[g_A^2(\Delta-\omega_L)+\omega_L\left(-g_D^2+(\delta-\omega_L)(\Delta-\omega_L)\right)\right]^2}
\end{equation}
\noindent Coherent pump analysis reveals that optimal energy transfer under such monochromatic drive is insured around the polariton frequencies with maximized values when pumping is directed at the middle polariton (as shown in Fig.~\ref{fig4}e). The analytical result listed above agrees very well with the numerical simulation used in Fig.~\ref{fig4}e. While in the incoherent pumping scenario, reaching the optimal transfer rate does not guarantee good concurrence, direct pumping into the middle polariton with monochromatic drive can give rise to entanglement (see Fig.~\ref{fig4}f).

\section*{Discussions and outlook}

\noindent The simplified model analyzed here could be relevant for experiments aiming at strong coupling of pairs of molecules to microcavity modes under low temperature conditions~\cite{wang2017coherent} and fulfilling the condition that donor-acceptor detunings as well as cavity linewidth are small compared to vibrational energies. This allowed us to focus on the zero-phonon line and to include extra decay channels only in the ground state vibrational manifold, justifying a three level quantum emitter approach that can easily be mapped onto a two-level effective model. More generally, in realistic scenarios, the donor and acceptor detunings can be large compared to vibrational energies which means that the vibrational manifold of the excited energy state (of the acceptor) has to be accounted for. A direct extension of our approach possibly allowing for an analytical understanding of the role of vibrations in the process of energy transfer inside cavities would assume the addition of localized molecular vibrations as $H_v=\nu_D b^\dagger_D b_D+\nu_A b^\dagger_A b_A$ at frequencies $\nu_{D,A}$ and annihilation operators $b_{D,A}$. The coupling between vibrations and electronic degrees of freedom can be simply modelled as a Holstein interaction Hamiltonian:
\begin{equation}
H_{e-v}=\lambda_D\nu_D(b^{\dagger}_D+b_D)\sigma^{\dagger}_D\sigma_D+\lambda_A\nu_A(b^{\dagger}_A+b_A)\sigma^{\dagger}_A\sigma_A.
\end{equation}
where the $\lambda_{D,A}$ characterize the mismatch between the equilibrium nuclear coordinates in the excited and ground states. An analytical treatment will be tackled in the future employing the diagonalizing polaron transformation applied to both donor and acceptor systems $\exp(\lambda (b-b^{\dagger})\sigma^{\dagger}\sigma)$. The expected result in the cavity case is that the bare cavity-donor-acceptor polaritons which are orthogonal in the absence of localized vibrations will couple. This leads to an additional transfer channel via vibrations-assisted polariton-polariton dissipative coupling. This would for example lead to a decay of an excitation introduced coherently in the upper polariton down to the lower polariton and therefore a considerable contribution to the transfer process.\\
As an illustration of possible follow-ups on this approach meant on increasing the complexity and relevance of our model we will consider sets of vibrational states of the acceptor denoted by a single state vector $\ket{i_A}$ close to the energy of the excited donor state. We can immediately supplement our model to include the dynamics of the intermediate states. Two possible paths for energy transfer occur now: i) direct off-resonant transfer from the donor to the acceptor low lying vibrational levels followed by emission and ii) resonant transfer from the donor to the intermediate states with high vibrational occupancy followed by non-radiative transfer and emission. While for $\Delta\gg\omega_v$ (with $\omega_v$ being the frequency of the vibrational level) the first process is negligible, the second process gives rise to rate equations marked by the common signature of the Förster resonant energy transfer mechanism:
\begin{eqnarray}
   \dot{p}_D &=&-(\Gamma+\gamma+\kappa_{FS}+\kappa_{ET})p_D+\Gamma,\\
   \dot{p}_A &=&-(\Gamma+\gamma)p_A+(\kappa_{FS}+\kappa_{ET})p_D,
\end{eqnarray}
showing transfer rates $\kappa_{FS}$ (for free space) and $\kappa_{ET}$ (for cavity) that also can be deduced as an increased linewidth of the donor.
indicating the typical signature of unidirectionality of the resonant F\"{o}rster energy transfer mechanism where population migrates from the donor to the acceptor and the modification of the natural linewidth of the donor occurs. In the low excitation regime where the two emitters behave linearly, such a model could be sufficient for simulating the behavior of molecular layers under collective strong coupling conditions~\cite{zhong2016non,zhong2017energy}.
While such a slightly more complex model to describe this dynamics will be tackled in the future, we point out here a first prediction of this result applicable in the free space scenario showing that the intermediate level flow rate is given by $J_i/\Gamma=4\Omega^2/(\gamma_D \gamma_{nr}+4\Omega^2)$ and dominates over the off-resonant transfer characterized by $\Delta^2$ in the denominator under the condition $\gamma_{nr}\ll \Delta$. For such unidirectional flow a energy transfer rate can be defined: $k_{FS}=4\Omega^2/\gamma_{nr}$. We will extend our analysis in future endeavors to tackle such a more complex model under simplified assumptions including coupling of electronic levels to vibrational and phononic baths.\\

\section*{Methods}

\subsection*{Vacuum mediated dipole-dipole coherent and incoherent interactions. Subradiance and superradiance.}

The complete expression of the dipole-dipole energy shift arising between two dipoles $\boldsymbol{\mu}_D$ and $\boldsymbol{\mu}_A$ separated by distance $ \mathbf{d}$ is given by

\begin{eqnarray}
   \Omega =\frac{3\sqrt{\gamma_D \gamma_A}}{2} \left\{-\left[ \left(\boldsymbol{\mu}_D \cdot \boldsymbol{\mu}_A\right)-\left(\boldsymbol{\mu}_D \cdot \mathbf{d}\right)\left(\boldsymbol{\mu}_A \cdot \mathbf{d}\right)\right] \frac {\cos k d} {k d}+\left[ \left(\boldsymbol{\mu}_D \cdot \boldsymbol{\mu}_A\right)-3\left(\boldsymbol{\mu}_D \cdot \mathbf{d}\right)\left(\boldsymbol{\mu}_A \cdot \mathbf{d}\right)\right]\left[ \frac {\sin k d} {(k d)^2}+ \frac {\cos k d} {(k d)^3}\right]\right\}.
\end{eqnarray}

\noindent The incoherent coupling that leads to collective decay is signaled by the mutual decay function:

\begin{eqnarray}
 \bar \gamma =\frac{3\sqrt{\gamma_D \gamma_A}}{2} \left\{\left[ \left(\boldsymbol{\mu}_D \cdot \boldsymbol{\mu}_A\right)-\left(\boldsymbol{\mu}_D \cdot \mathbf{d}\right)\left(\boldsymbol{\mu}_A \cdot \mathbf{d}\right)\right] \frac {\sin k d} {k d}+ \left[ \left(\boldsymbol{\mu}_D \cdot \boldsymbol{\mu}_A\right)-3\left(\boldsymbol{\mu}_D \cdot \mathbf{d}\right)\left(\boldsymbol{\mu}_A \cdot \mathbf{d}\right)\right]\left[ \frac {\cos k d} {(k d)^2}- \frac {\sin k d} {(k d)^3}\right]\right\}.
\end{eqnarray}

\noindent In the main text we used simplified expressions for the above quantities by assuming $\boldsymbol{\mu}_A=\boldsymbol{\mu}_D$ so that they are always parallel to each other and only vary the angle between the dipoles and the inter-particle separation direction. To signal the occurrence of subradiance and superradiance, let us consider two quantum emitters separated by $d$ with a dipole-dipole interaction strength $\Omega(d)$. Assuming equal transition frequencies of the donor and acceptor, diagonalization of the Hamiltonian in the presence of the interaction leads to new eigenstates $1/\sqrt{2}(|e_Dg_A\rangle \pm |g_De_A\rangle )$. Rewriting the Lindblad terms one finds that two decay channels are present, one for the symmetric combination $1/\sqrt{2}(|e_Dg_A\rangle + |g_De_A\rangle )$ with decay rate $\gamma+\bar{\gamma}$ and one for the asymmetric combination $1/\sqrt{2}(|e_Dg_A\rangle - |g_De_A\rangle )$ with decay rate $\gamma-\bar{\gamma}$. In the limit of very small separations, the mutual decay rate tends to $\gamma$ such that the symmetric state becomes fully superradiant decaying at the rate $2\gamma$ while the asymmetric state becomes fully subradiant at effective zero decay rate.

\subsection*{Model for incoherent pumping}

Let us assume three levels of interest in the donor: excited, ground and some intermediate level that can very quickly decay onto the excited state at rate $\gamma_{ie}$. Driving from the ground to the intermediate level takes place at the Rabi frequency  and is followed by quick relaxation into the excited state with $\gamma_{ie}$. The excited state population as well as the $i-g$ coherence are therefore vanishingly small and one can derive an effective two-level model. To derive this model let us write the Bloch equations for the $i-e-g$ system:
\begin{subequations}
\begin{align}
\frac{d}{dt}{\rho_{ee}}&=-\gamma_e \rho_{ee}+\gamma_{ie} \rho_{ii},\\
\frac{d}{dt}{\rho_{eg}}&= -\frac{\gamma_{eg}}{2}\rho_{eg} +i \eta \rho_{ei}, \\
\frac{d}{dt}{\rho_{gg}}&=\gamma_e \rho_{ee}+\gamma_{ig} \rho_{ii}+i \eta (\rho_{gi}-\rho_{ig}), \\
\frac{d}{dt}{\rho_{ii}}&=-\gamma_{ig} \rho_{ii}-\gamma_{ie}\rho_{ii}-i \eta (\rho_{gi}-\rho_{ig}), \\
\frac{d}{dt}{\rho_{ig}}&=-\frac{\gamma_{ie}+\gamma_{ig}}{2}\rho_{ig}-i \eta (\rho_{gg}-\rho_{ii}), \\
\frac{d}{dt}{\rho_{ei}}&=-\frac{\gamma_{ie}+\gamma_{ig}+\gamma_{eg}}{2}\rho_{ei}+i \eta \rho_{eg}.
\end{align}
\end{subequations}
Elimination of the fastly decaying elements $\rho_{ig}$ and $\rho_{ii}$ and $\rho_{ei}$ leads to a set of coupled equations for a complete two level system:
\begin{subequations}
\begin{align}
\frac{d}{dt}{\rho_{ee}}&=-\gamma_e \rho_{ee}+\frac{4\eta^2 \gamma_{ie}}{4\eta^2+(\gamma_{ie}+\gamma_{ig})^2} \rho_{gg},\\
\frac{d}{dt}{\rho_{eg}}&= -\frac{\gamma_{eg}}{2}\rho_{eg} - \frac{2\eta^2}{\gamma_{ie}+\gamma_{ig}+\gamma_{eg}} \rho_{eg}, \\
\frac{d}{dt}{\rho_{gg}}&=\gamma_e \rho_{ee}-\frac{4\eta^2 \gamma_{ie}}{4\eta^2+(\gamma_{ie}+\gamma_{ig})^2} \rho_{gg}, \\
\end{align}
\end{subequations}
In the limit that $\gamma_{ie}\gg \gamma_{eg},\gamma_{ig},\eta$ one can approximate all the new rates with an effective pump rate $\Gamma=4\eta^2/\gamma_{ie}$. The dynamics is then the opposite of the spontaneous decay, namely a Lindblad term with an $|e\rangle \langle g|$ collapse operator.

\subsection*{Equations of motion for operators}
Starting with the master equation one can derive equations of motion for single operator averages:
\begin{subequations}
\label{motion}
\begin{align}
\frac{d}{dt}{\braket{a}}&=\left(-i\delta-\frac{\kappa}{2}\right)\braket{a}-ig_D\braket{\sigma_D}-ig_A\braket{\sigma_A},\\\frac{d}{dt}{\braket{\sigma_D}}&=\left(-i\Delta-\frac{\gamma_{tot}}{2}-\frac{\Gamma}{2}\right)\braket{\sigma_D}-i\Omega\braket{\sigma_A}-ig_D\braket{a}-\frac{\bar{\gamma}}{2}\braket{\sigma_A}+\bar{\gamma}\braket{\hat{p}_D\sigma_A}, \\
\frac{d}{dt}{\braket{\sigma_A}}&=\left(-\frac{\gamma_{tot}}{2}-\frac{\Gamma}{2}\right)\braket{\sigma_A}-i\Omega\braket{\sigma_D}-ig_A\braket{a}-\frac{\bar{\gamma}}{2}\braket{\sigma_D}+\bar{\gamma}\braket{\hat{p}_A\sigma_D}.
\end{align}
\end{subequations}
While the equations are not linear as higher order correlations are present as source terms, under the assumption of weak pumping, terms containing products of population with coherence operators will be neglected. Notice also that when one sets $g_A=g_D=0$ one can simply diagonalize the linearized form of the equations above to derive the collective shift and decay rates of symmetric (superradiant) and asymmetric (subradiant) states.

\subsection*{Complete set of equations of motion for two-body correlations}

For the derivation of quantities of interest such as energy flow rates and populations in steady state we start with the full master equation and compute equations of motion for operator-operator correlations:
\begin{subequations}
\label{complete}
\begin{eqnarray}
\frac{d}{dt}{\braket{\sigma_D^\dagger\sigma_D}}=\dot{p}_D=\Gamma-(\Gamma+\gamma_{tot})p_D-J-J_D-\bar{\gamma}\frac{J^r}{2\Omega}, \\
\frac{d}{dt}{\braket{\sigma_A^\dagger\sigma_A}}=\dot{p}_A=-(\Gamma+\gamma_{tot})p_A+J+J_A-\bar{\gamma}\frac{J^r}{2\Omega} , \\
\frac{d}{dt}{\braket{a^\dagger a}}=\dot{n}=-\kappa n +J_D-J_A,  \\
\begin{split}
\frac{d}{dt}{\braket{\sigma_D^\dagger \sigma_A}}=i\Omega(p_A-p_D)-(\Gamma+\gamma_{tot}+2\gamma_\phi-i\Delta)\braket{\sigma_D^\dagger \sigma_A} -2ig_D\braket{a^\dagger \hat{p}_D\sigma_A}+2ig_A\braket{a\hat{p}_A\sigma_D^\dagger} \\ +ig_D\braket{a^\dagger\sigma_A}+ig_A\braket{a^\dagger \sigma_D}+2\bar{\gamma}\braket{\hat{p}_A\hat{p}_D}-\frac{\bar{\gamma}}{2}(p_A+p_D), \end{split}  \\
\begin{split}
\frac{d}{dt}{\braket{a^\dagger \sigma_D}}=ig_D(p_D-n)-\left[\frac{\Gamma+\gamma_{tot}+2\gamma_\phi+\kappa}{2}-i(\delta-\Delta)\right]\braket{a^\dagger \sigma_D} +2ig_D\braket{\hat{n}\hat{p}_D}+2i\Omega\braket{a^\dagger \hat{p}_D\sigma_A} \\-i\Omega\braket{a^\dagger\sigma_A}-ig_A\braket{\sigma_D^\dagger \sigma_A}+\bar{\gamma}\braket{a^\dagger\hat{p}_D\sigma_A}-\frac{\bar{\gamma}}{2}\braket{a^\dagger\sigma_A}, \end{split} \\
\begin{split}
\frac{d}{dt}{\braket{a^\dagger \sigma_A}}=ig_A (p_A-n)-\left[\frac{\Gamma+\gamma_{tot}+2\gamma_\phi+\kappa}{2}-i\delta\right]\braket{a^\dagger \sigma_A}+2ig_A\braket{\hat{n}\hat{p}_A}+2i\Omega\braket{a^\dagger \hat{p}_A\sigma_D} \\-i\Omega\braket{a^\dagger \sigma_D}+ig_D\braket{\sigma_D^\dagger\sigma_A}+\bar{\gamma}\braket{a^\dagger\hat{p}_A\sigma_D}-\frac{\bar{\gamma}}{2}\braket{a^\dagger\sigma_D}. \end{split}
\label{end}
\end{eqnarray}
\end{subequations}
We have defined $J^r=2\Omega\Re\braket{\sigma_D \sigma_A^\dagger}$. Solutions in the particular cases analyzed throughout the manuscript are obtained by setting the derivatives to zero which corresponds to steady state conditions. The existence of steady state solutions is validated by an analysis of the real parts of the drift matrix. For $g_A, g_D=0$ and assuming weak pumping $\Gamma\ll\gamma$ we obtain
\begin{equation}
J\propto \frac{2\Omega^2\gamma_{tot}}{\gamma_{tot}^2+\Delta^2}(p_D-p_A),
\end{equation}
signaling the dependence of $J$ on the population gradient between donor and acceptor.

\subsection*{Adiabatic elimination of the field in the bad cavity regime}

The equation of motion for the cavity field amplitude reads
\begin{equation}
\frac{d}{d t}\braket{a(t)}=\left(-i\delta-\frac{\kappa}{2}\right)\braket{a(t)}-ig_D\braket{\sigma_D(t)}-ig_A\braket{\sigma_A(t)}.
\label{eomcavity}
\end{equation}
A formal integration leads to the following solution
\begin{equation}
\braket{a(t)}=\braket{a(0)}e^{-\left(\kappa/2+i\delta\right)t}-ig_D\int_0^t\braket{\sigma_D\left(t'\right)}e^{-\left(\kappa/2+i\delta \right)(t-t')}dt'-i g_A \int_0^t \braket{\sigma_A (t')} e^{-\left(\kappa/2+i\delta\right)(t-t')}dt',
\end{equation}
which can be approximated in the bad cavity regime where $\kappa$ is larger than any other parameter. To this end, we assume evolution of the coherence at their natural frequencies and integrate the exponents to arrive at the expression in Eq.~\ref{adelim}. Replacing $\braket{a(t)}$ with the adiabatic elimination result in Eqs.~\ref{motion} we find the cavity mediated effective decay rates in Eqs.~\ref{effdecay}.

\subsection*{Concurrence}
After tracing out the light mode from the density operator of donor-acceptor-cavity field one obtains a two qubit mixed state described by a total density matrix $\rho$; as a measure of entanglement the concurrence is defined as
\begin{equation}
C(\rho)=\max(0, \lambda_1-\lambda_2-\lambda_3-\lambda_4),
\end{equation}
where $\lambda_i$ are the eigenvalues, in decreasing order, of the Hermitian operator
\begin{equation}
R=\sqrt{\sqrt{\rho}\tilde{\rho}\sqrt{\rho}}
\end{equation}
and
\begin{equation}
\tilde{\rho}=\left(\sigma_y \otimes \sigma_y \right)\rho^*\left(\sigma_y \otimes \sigma_y \right).
\end{equation}
Our procedure is numerical~\cite{qutip2013} consisting of a partial trace over the cavity mode degree of freedom to derive the reduced $\rho_{DA}$ and applying the steps above to the reduced matrix.\\

\bibliography{et}

\begin{thebibliography}{10}
\expandafter\ifx\csname url\endcsname\relax
  \def\url#1{\texttt{#1}}\fi
\expandafter\ifx\csname urlprefix\endcsname\relax\def\urlprefix{URL }\fi
\providecommand{\bibinfo}[2]{#2}
\providecommand{\eprint}[2][]{\url{#2}}

\bibitem{thomas1978energy}
\bibinfo{author}{Thomas, D.~D.}, \bibinfo{author}{Carlsen, W.~F.} \&
  \bibinfo{author}{Stryer, L.}
\newblock \bibinfo{title}{Fluorescence energy transfer in the rapid-diffusion
  limit}.
\newblock \emph{\bibinfo{journal}{Proceedings of the National Academy of
  Sciences of the United States of America}} \textbf{\bibinfo{volume}{75}},
  \bibinfo{pages}{5746--5750} (\bibinfo{year}{1978}).
\newblock \urlprefix\url{http://www.jstor.org/stable/68832}.

\bibitem{scholes2003}
\bibinfo{author}{Scholes, G.~D.}
\newblock \bibinfo{title}{Long-range resonance transfer in molecular systems}.
\newblock \emph{\bibinfo{journal}{Annual Review of Physical Chemistry}}
  \textbf{\bibinfo{volume}{54}}, \bibinfo{pages}{57--87}
  (\bibinfo{year}{2003}).
\newblock
  \urlprefix\url{https://doi.org/10.1146/annurev.physchem.54.011002.103746}.
\newblock \bibinfo{note}{PMID: 12471171}.

\bibitem{loura2012simple}
\bibinfo{author}{Loura, L. M.~S.}
\newblock \bibinfo{title}{Simple estimation of förster resonance energy
  transfer (fret) orientation factor distribution in membranes}.
\newblock \emph{\bibinfo{journal}{International Journal of Molecular Sciences}}
  \textbf{\bibinfo{volume}{13}}, \bibinfo{pages}{15252} (\bibinfo{year}{2012}).
\newblock
  \urlprefix\url{https://www.ncbi.nlm.nih.gov/pmc/articles/PMC3509639/}.

\bibitem{wang2017coherent}
\bibinfo{author}{Wang, D.} \emph{et~al.}
\newblock \bibinfo{title}{Coherent coupling of a single molecule to a scanning
  fabry-perot microcavity}.
\newblock \emph{\bibinfo{journal}{Phys. Rev. X}} \textbf{\bibinfo{volume}{7}},
  \bibinfo{pages}{021014} (\bibinfo{year}{2017}).
\newblock \urlprefix\url{https://link.aps.org/doi/10.1103/PhysRevX.7.021014}.

\bibitem{zhong2016non}
\bibinfo{author}{Zhong, X.} \emph{et~al.}
\newblock \bibinfo{title}{Non-radiative energy transfer mediated by hybrid
  light-matter states}.
\newblock \emph{\bibinfo{journal}{Angewandte Chemie}}
  \textbf{\bibinfo{volume}{55}}, \bibinfo{pages}{6202} (\bibinfo{year}{2016}).
\newblock
  \urlprefix\url{http://onlinelibrary.wiley.com/doi/10.1002/anie.201600428/fulll}.

\bibitem{zhong2017energy}
\bibinfo{author}{Zhong, X.} \emph{et~al.}
\newblock \bibinfo{title}{Energy transfer between spatially separated entangled
  molecules}.
\newblock \emph{\bibinfo{journal}{Angewandte Chemie}}
  \textbf{\bibinfo{volume}{56}}, \bibinfo{pages}{9034} (\bibinfo{year}{2017}).
\newblock
  \urlprefix\url{http://onlinelibrary.wiley.com/doi/10.1002/anie.201703539/full}.

\bibitem{orgiu2015conductivity}
\bibinfo{author}{Orgiu, E.} \emph{et~al.}
\newblock \bibinfo{title}{Conductivity in organic semiconductors hybridized
  with the vacuum field}.
\newblock \emph{\bibinfo{journal}{Nature Materials}}
  \textbf{\bibinfo{volume}{14}}, \bibinfo{pages}{1123 -- 1129}
  (\bibinfo{year}{2015}).
\newblock
  \urlprefix\url{http://www.nature.com/nmat/journal/vaop/ncurrent/full/nmat4392.html}.

\bibitem{feist2015extraordinary}
\bibinfo{author}{Feist, J.} \& \bibinfo{author}{Garcia-Vidal, F.~J.}
\newblock \bibinfo{title}{Extraordinary exciton conductance induced by strong
  coupling}.
\newblock \emph{\bibinfo{journal}{Phys. Rev. Lett.}}
  \textbf{\bibinfo{volume}{114}}, \bibinfo{pages}{196402}
  (\bibinfo{year}{2015}).
\newblock
  \urlprefix\url{http://link.aps.org/doi/10.1103/PhysRevLett.114.196402}.

\bibitem{schachenmayer2015cavity}
\bibinfo{author}{Schachenmayer, J.}, \bibinfo{author}{Genes, C.},
  \bibinfo{author}{Tignone, E.} \& \bibinfo{author}{Pupillo, G.}
\newblock \bibinfo{title}{Cavity-enhanced transport of excitons}.
\newblock \emph{\bibinfo{journal}{Phys. Rev. Lett.}}
  \textbf{\bibinfo{volume}{114}}, \bibinfo{pages}{196403}
  (\bibinfo{year}{2015}).
\newblock
  \urlprefix\url{http://link.aps.org/doi/10.1103/PhysRevLett.114.196403}.

\bibitem{hagenmuller2017cavity}
\bibinfo{author}{Hagenm\"uller, D.}, \bibinfo{author}{Schachenmayer, J.},
  \bibinfo{author}{Sch\"utz, S.}, \bibinfo{author}{Genes, C.} \&
  \bibinfo{author}{Pupillo, G.}
\newblock \bibinfo{title}{Cavity-enhanced transport of charge}.
\newblock \emph{\bibinfo{journal}{Phys. Rev. Lett.}}
  \textbf{\bibinfo{volume}{119}}, \bibinfo{pages}{223601}
  (\bibinfo{year}{2017}).
\newblock
  \urlprefix\url{https://link.aps.org/doi/10.1103/PhysRevLett.119.223601}.

\bibitem{hagenmuller2018fermions}
\bibinfo{author}{Hagenm\"uller, D.}, \bibinfo{author}{Sch\"utz, S.},
  \bibinfo{author}{Schachenmayer, J.}, \bibinfo{author}{Genes, C.} \&
  \bibinfo{author}{Pupillo, G.}
\newblock \bibinfo{title}{Cavity-assisted mesoscopic transport of fermions:
  Coherent and dissipative dynamics}.
\newblock \emph{\bibinfo{journal}{arxiv:180109876}}  (\bibinfo{year}{2018}).
\newblock \urlprefix\url{https://arxiv.org/abs/1801.09876}.

\bibitem{shalabney2015coherent}
\bibinfo{author}{Shalabney, A.} \emph{et~al.}
\newblock \bibinfo{title}{Coherent coupling of molecular resonators with a
  microcavity mode}.
\newblock \emph{\bibinfo{journal}{Nature Communications}}
  \textbf{\bibinfo{volume}{6}}, \bibinfo{pages}{5981} (\bibinfo{year}{2015}).
\newblock \urlprefix\url{http://dx.doi.org/10.1038/ncomms6981}.

\bibitem{george2015liquid}
\bibinfo{author}{George, J.}, \bibinfo{author}{Shalabney, A.},
  \bibinfo{author}{Hutchison, J.~A.}, \bibinfo{author}{Genet, C.} \&
  \bibinfo{author}{Ebbesen, T.~W.}
\newblock \bibinfo{title}{Liquid-phase vibrational strong coupling}.
\newblock \emph{\bibinfo{journal}{The Journal of Physical Chemistry Letters}}
  \textbf{\bibinfo{volume}{6}}, \bibinfo{pages}{1027--1031}
  (\bibinfo{year}{2015}).
\newblock \urlprefix\url{http://dx.doi.org/10.1021/acs.jpclett.5b00204}.

\bibitem{galego2016suppressing}
\bibinfo{author}{J.~Galego, F. G.-V.} \& \bibinfo{author}{Feist, J.}
\newblock \bibinfo{title}{Suppressing photochemical reactions with quantized
  light fields}.
\newblock \emph{\bibinfo{journal}{Nature Communications}}
  \bibinfo{pages}{13841} (\bibinfo{year}{2017}).
\newblock \urlprefix\url{https://www.nature.com/articles/ncomms13841}.

\bibitem{galego2017many}
\bibinfo{author}{Galego, J.}, \bibinfo{author}{Garcia-Vidal, F.~J.} \&
  \bibinfo{author}{Feist, J.}
\newblock \bibinfo{title}{Many-molecule reaction triggered by a single photon
  in polaritonic chemistry}.
\newblock \emph{\bibinfo{journal}{Phys. Rev. Lett.}}
  \textbf{\bibinfo{volume}{119}}, \bibinfo{pages}{136001}
  (\bibinfo{year}{2017}).
\newblock
  \urlprefix\url{https://link.aps.org/doi/10.1103/PhysRevLett.119.136001}.

\bibitem{flick2017atoms}
\bibinfo{author}{Flick, J.}, \bibinfo{author}{Ruggenthaler, M.},
  \bibinfo{author}{Appel, H.} \& \bibinfo{author}{Rubio, A.}
\newblock \bibinfo{title}{Atoms and molecules in cavities, from weak to strong
  coupling in quantum-electrodynamics (qed) chemistry}
  \textbf{\bibinfo{volume}{114}}, \bibinfo{pages}{3026--3034}
  (\bibinfo{year}{2017}).

\bibitem{flick2017cavity}
\bibinfo{author}{Flick, J.}, \bibinfo{author}{Appel, H.},
  \bibinfo{author}{Ruggenthaler, M.} \& \bibinfo{author}{Rubio, A.}
\newblock \bibinfo{title}{Cavity born–oppenheimer approximation for
  correlated electron–nuclear-photon systems}.
\newblock \emph{\bibinfo{journal}{Journal of Chemical Theory and Computation}}
  \textbf{\bibinfo{volume}{13}}, \bibinfo{pages}{1616--1625}
  (\bibinfo{year}{2017}).
\newblock \urlprefix\url{https://doi.org/10.1021/acs.jctc.6b01126}.
\newblock \bibinfo{note}{PMID: 28277664},
  \eprint{https://doi.org/10.1021/acs.jctc.6b01126}.

\bibitem{coles2013polariton}
\bibinfo{author}{Coles, D.~M.} \emph{et~al.}
\newblock \bibinfo{title}{Polariton-mediated energy transfer between organic
  dyes in a strongly coupled optical microcavity}.
\newblock \emph{\bibinfo{journal}{Nature Materials}}
  \textbf{\bibinfo{volume}{13}}, \bibinfo{pages}{712} (\bibinfo{year}{2013}).
\newblock \urlprefix\url{https://www.nature.com/articles/nmat3950}.

\bibitem{feist2017long}
\bibinfo{author}{Garcia-Vidal, F.~J.} \& \bibinfo{author}{Feist, J.}
\newblock \bibinfo{title}{Long-distance operator for energy transfer}.
\newblock \emph{\bibinfo{journal}{Science}} \textbf{\bibinfo{volume}{357}},
  \bibinfo{pages}{1357} (\bibinfo{year}{2017}).
\newblock \urlprefix\url{http://science.sciencemag.org/content/357/6358/1357}.

\bibitem{vidal2018organic}
\bibinfo{author}{Sáez-Blázquez, R.}, \bibinfo{author}{Feist, J.},
  \bibinfo{author}{Fernández-Domínguez, A.~I.} \&
  \bibinfo{author}{García-Vidal, F.~J.}
\newblock \bibinfo{title}{Organic polaritons enable local vibrations to drive
  long-range energy transfer}.
\newblock \emph{\bibinfo{journal}{arxiv:1804.01784v1}}  (\bibinfo{year}{2018}).
\newblock \urlprefix\url{https://arxiv.org/abs/1804.01784}.

\bibitem{joel2018theory}
\bibinfo{author}{Du, M.} \emph{et~al.}
\newblock \bibinfo{title}{Theory for polariton-assisted remote energy
  transfer}.
\newblock \emph{\bibinfo{journal}{arXiv:1711.11576}}  (\bibinfo{year}{2018}).
\newblock \urlprefix\url{https://arxiv.org/abs/1711.11576}.

\bibitem{gotzinger2006controlled}
\bibinfo{author}{G\"{o}tzinger, S.} \emph{et~al.}
\newblock \bibinfo{title}{Controlled photon transfer between two individual
  nanoemitters via shared high-q modes of a microsphere resonator}.
\newblock \emph{\bibinfo{journal}{Nano Letters}} \textbf{\bibinfo{volume}{6}},
  \bibinfo{pages}{1151} (\bibinfo{year}{2006}).
\newblock \urlprefix\url{http://pubs.acs.org/doi/abs/10.1021/nl060306p}.

\bibitem{cano2010resonance}
\bibinfo{author}{Martín-Cano, D.}, \bibinfo{author}{Martín-Moreno, L.},
  \bibinfo{author}{García-Vidal, F.~J.} \& \bibinfo{author}{Moreno, E.}
\newblock \bibinfo{title}{Resonance energy transfer and superradiance mediated
  by plasmonic nanowaveguides}.
\newblock \emph{\bibinfo{journal}{Nano Letters}} \textbf{\bibinfo{volume}{10}},
  \bibinfo{pages}{3129--3134} (\bibinfo{year}{2010}).
\newblock \urlprefix\url{http://dx.doi.org/10.1021/nl101876f}.

\bibitem{Duque2015quantumclassical}
\bibinfo{author}{Duque, S.}, \bibinfo{author}{Brumer, P.} \&
  \bibinfo{author}{Pach\'on, L.~A.}
\newblock \bibinfo{title}{Classical approach to multichromophoric resonance
  energy transfer}.
\newblock \emph{\bibinfo{journal}{Phys. Rev. Lett.}}
  \textbf{\bibinfo{volume}{115}}, \bibinfo{pages}{110402}
  (\bibinfo{year}{2015}).
\newblock
  \urlprefix\url{https://link.aps.org/doi/10.1103/PhysRevLett.115.110402}.

\bibitem{ficek1987quantum}
\bibinfo{author}{Ficek, Z.}, \bibinfo{author}{Tanaś, R.} \&
  \bibinfo{author}{Kielich, S.}
\newblock \bibinfo{title}{Quantum beats and superradiant effects in the
  spontaneous emission from two nonidentical atoms}.
\newblock \emph{\bibinfo{journal}{Physica A: Statistical Mechanics and its
  Applications}} \textbf{\bibinfo{volume}{146}}, \bibinfo{pages}{452 -- 482}
  (\bibinfo{year}{1987}).
\newblock
  \urlprefix\url{http://www.sciencedirect.com/science/article/pii/0378437187902809}.

\bibitem{Pachon2017incoherent}
\bibinfo{author}{Pachón, L.~A.}, \bibinfo{author}{Botero, J.~D.} \&
  \bibinfo{author}{Brumer, P.}
\newblock \bibinfo{title}{Open system perspective on incoherent excitation of
  light-harvesting systems}.
\newblock \emph{\bibinfo{journal}{Journal of Physics B: Atomic, Molecular and
  Optical Physics}} \textbf{\bibinfo{volume}{50}}, \bibinfo{pages}{184003}
  (\bibinfo{year}{2017}).
\newblock \urlprefix\url{http://stacks.iop.org/0953-4075/50/i=18/a=184003}.

\bibitem{niemczyk2010ultrastrong}
\bibinfo{author}{Niemczyk, T.} \emph{et~al.}
\newblock \bibinfo{title}{Circuit quantum electrodynamics in the
  ultrastrong-coupling regime}.
\newblock \emph{\bibinfo{journal}{Nature Physics}}
  \textbf{\bibinfo{volume}{6}}, \bibinfo{pages}{772--776}
  (\bibinfo{year}{2010}).
\newblock \urlprefix\url{http://www.nature.com/articles/nphys1730}.

\bibitem{qutip2013}
\bibinfo{author}{Johansson, J.~R.}, \bibinfo{author}{Nation, P.~D.} \&
  \bibinfo{author}{Nori, F.}
\newblock \bibinfo{title}{Qutip 2: A python framework for the dynamics of open
  quantum systems}.
\newblock \emph{\bibinfo{journal}{Comp. Phys. Comm.}}
  \textbf{\bibinfo{volume}{184}}, \bibinfo{pages}{1234--1240}
  (\bibinfo{year}{2013}).
\newblock \urlprefix\url{http://www.sciencedirect.com/science/article/pii/
  S0010465512003955}.

\end{thebibliography}

\section*{Acknowledgements}

We acknowledge useful disscusions with T.~W.~Ebbesen, V.~Sandoghdar, S.~G\"{o}tzinger, D.~Wang, H.~Kelkar, D.~Martin-Cano, G.~Pupillo, D.~Hagenmuller, C.~Genet and J.~Schachenmayer. Financial support has been provided by the Max Planck Society.

\section*{Author contributions statement}
M.~R. has been the main investigator in this project taking part in developing the model and goals, performing analytical calculations, numerical simulations, producing figures and taking big part in writing. F.~M. performed analytical calculations in the strong coupling regime and numerical simulations. C.~G. developed the ideas, participated in calculations and simulations, supervised the project and took lead in the writing process.

\section*{Additional information}

We declare no competing interests and no financial interests arise in the funding of this work.

\end{document}